\documentclass[aps,prd,a4paper,reprint,nofootinbib,superscriptaddress,floatfix,preprintnumbers]{revtex4-2}

\usepackage{xurl}
\usepackage{hyperref}
\usepackage{amsmath}
\usepackage{amsfonts}
\usepackage{amssymb}
\usepackage{color}
\usepackage[utf8]{inputenc}
\usepackage{graphicx}
\usepackage{microtype}
\usepackage{siunitx}
\usepackage{soul}
\usepackage{longtable}
\usepackage{bm}
\usepackage{xspace}
\usepackage{natbib}
\usepackage[normalem]{ulem}
\usepackage[table]{xcolor}
\usepackage{comment}

\renewcommand{\arraystretch}{1.5}

\usepackage{lineno}

\begin{document}

\begin{flushleft}
LAPTH-043/25
\end{flushleft}

\title{Low-frequency radio telescopes sensitivity to light dark matter}
\date{\today}
\author{Ruben Zatini}\email{zatini@lapth.cnrs.fr}\email{rzatini@iac.es}
\affiliation{LAPTh, CNRS,  USMB, F-74940 Annecy, France}
\affiliation{Instituto de Astrof\'isica de Canarias,  C/ V\'ia L\'actea, s/n E38205 - La Laguna, Tenerife, Spain}
\affiliation{Universidad de La Laguna, Departamento de Astrof\'isica, La Laguna, Tenerife, Spain}
\author{Francesca Calore}
\affiliation{LAPTh, CNRS,  USMB, F-74940 Annecy, France}
\author{Pasquale Dario Serpico}
\affiliation{LAPTh, CNRS,  USMB, F-74940 Annecy, France}

\begin{abstract}
Ground-based radio telescopes are routinely used to search for light dark matter (DM) candidates such as axion-like particles or dark photons. 
However, these instruments face inherent limitations to push the searches to masses below $10^{-7}$ eV, due to the effect of the Earth's ionosphere. The extant and planned space- or Moon-based radio telescopes motivate this study: We systematically investigate their sensitivity to resonant conversion of  light DM into radio signals from three solar system targets: the Sun, the Earth, and Jupiter. The perspectives are especially encouraging for dark photon searches using the Sun 
as a target, and for axion-like particles conversion in  Jupiter's magnetosphere.
\end{abstract}

\maketitle

\section{Introduction}\label{sec:intro}
The search for light and (ultra-)weakly interacting  particles has emerged as an exciting frontier in particle physics, offering potential solutions to long-standing puzzles, like the nature of dark matter (DM). 
Some models are particularly compelling, such as the QCD axion,  which also addresses the strong CP problem~\cite{Peccei:1977hh,Peccei:1977ur,Weinberg:1977ma,Wilczek:1977pj}. A broader class of axion-like particles (ALPs) or vector fields dubbed dark photons (DPs) also make good DM candidates in some part of their parameter space (see e.g.~Ref.~\cite{Arias:2012az}).
In looking for these new states, laboratory-based searches and astrophysical observations have demonstrated remarkable complementarity, with both approaches being competitive~\cite{Giannotti:2024xhx,Fabbrichesi:2020wbt}.

The radio frequency window has received significant attention for probing the parameter space of such particles. 
A particularly promising avenue involves the conversion of DM in strongly magnetised objects such as neutron stars~\cite{Pshirkov:2007st, PhysRevD.97.123001,PhysRevD.101.123003,PhysRevLett.121.241102,PhysRevD.104.103030,Millar_2021,Battye_2021}. Additionally, the exploration of conversion processes in other astrophysical systems, such as the Sun --where signals could be detected both in space~\cite{2025PhRvL.134q1001A} and on the ground~\cite{2023Univ....9..142A,2024NatCo..15..915A,2024PhLB..85438752T}-- and the Earth~\cite{2024PhRvL.133y1001B,2023Symm...15.1167C}, further expands the scope of detection strategies.

However, ground-based radio telescopes face inherent limitations due to the Earth's ionosphere, which reflects or absorbs signals below approximately 10 MHz, thereby restricting observations to higher frequencies \cite{1977MNRAS.179...21C, 2020JAI.....950019C}. While some experiments have attempted to exploit favourable ionospheric conditions to extend measurements below this threshold, the challenges remain substantial.
In contrast, the current and next-generation space-based radio telescopes provide an alternative, as they are not constrained by the Earth's ionosphere and can access lower radio frequencies. The scientific case of these low-frequency radio instruments is rich, ranging from planetary science to cosmology, notably if a lunar observatory could be built~\cite{2009NewAR..53....1J}.
Moreover, their capabilities also open new opportunities for detecting ALPs and DPs, a fact that is more rarely acknowledged.

The potential advantages of space-based observations motivate us to systematically investigate the feasibility and sensitivity of low frequency searches for light DM.
In this work, we undertake a comparative study to assess the advantages of low radio frequency observations for ALPs and DPs searches. By exploring irreducible astrophysical backgrounds and the current and future experimental landscape, we aim to clarify the potential for probing these elusive particles.

The paper is structured as follows: In Sec.~\ref{sec:signals}, 
we introduce the theoretical framework of light DM conversion in astrophysical unmagnetised objects of the Solar system: the Sun, Jupiter, and the Earth. In Sec.~\ref{sec:detection}, we highlight the phenomenology of the DM signatures from the three targets, and we provide an overview on detectability of the signals, and characterisation of sky backgrounds and instruments' noise.
In Sec.~\ref{sec:results}, we present sensitivity projections for both DM ALPs and DPs for the different targets and instruments.  
We discuss our results and conclude in Sec.~\ref{sec:conclusions}.

\section{Dark matter-photon conversion}\label{sec:signals}
In this section, we describe the theoretical framework for DM conversion into photons in unmagnetised astrophysical targets, for both ALPs and DPs. 
All expressions that involve particle masses or  frequencies are most compactly written in natural units ($\hbar=c=1$).

\subsection{Resonant conversion probability}
The mixing between DM candidates (ALP and DP) and photons in a plasma can be described by a coupled DM--photon system, obtained through simple extensions of the quantum electrodynamics (QED) Lagrangian. For DPs, one introduces an additional hidden $U(1)$ gauge boson (described by the field $A_\mu'$ and tensor $F_{\mu\nu}' =\partial_\mu A_\nu'-\partial_\nu A_\mu'$) of mass $m_{A'}$ with a kinetic mixing term with the photon (described by the tensor $F_{\mu\nu}$)~\cite{Okun:1982xi,1986PhLB..166..196H}
\begin{equation}
    \begin{aligned}
        \mathcal{L}_{\rm QED+A'} \;\supset\;
        & -\dfrac{1}{4} F_{\mu\nu}F^{\mu\nu} 
        - \dfrac{1}{4} F'_{\mu\nu}F'^{\mu\nu} + \dfrac{\epsilon}{2} F_{\mu\nu}F'^{\mu\nu} \\
        &
        + \dfrac{1}{2} m_{A'}^2 A'_\mu A'^\mu ,
    \end{aligned}
\end{equation}
where the parameter $\epsilon$ governs the mixing. For the ALP field $a$ of mass $m_a$, the interaction arises from the ALP--photon coupling $g_{a\gamma\gamma}$~\cite{Sikivie:1983ip}, 
\begin{equation}
    \begin{aligned}
        \mathcal{L}_{\rm QED+a} \;\supset\;
        & -\dfrac{1}{4} F_{\mu\nu}F^{\mu\nu} 
        + \dfrac{1}{2} (\partial_\mu a)(\partial^\mu a)
        - \dfrac{1}{2} m_a^2 a^2 \\
        &
        - \dfrac{1}{4} g_{a\gamma\gamma}\, a\, F_{\mu\nu}\tilde{F}^{\mu\nu} .
    \end{aligned}
\end{equation}
In this case, an oscillating ALP field in the presence of an external magnetic field induces an effective current 
$\mathbf{J}_{\rm eff} = g_{a\gamma\gamma}\, \dot a\, \mathbf{B}_{\rm ext}$, which sources electromagnetic waves. In both cases, the system reduces to modified Maxwell equations describing the mixing of two fields (the DM candidate and the photon). 

The presence of a DM mass induces a momentum mismatch between the DM field and the photon, which suppresses the conversion between the two particles. However, in a plasma the photon acquires an effective mass given by the plasma frequency
\begin{equation}\label{eq:plasma_frequency}
\omega_{\rm pl} = \left( \dfrac{4 \pi \, \alpha_{\rm EM}\, n_e}{m_e}\right)^{\frac12},
\end{equation}
where $\alpha_{\rm EM}$ is the fine structure constant, $n_e$ is the local electron density and $m_e$ is the electron mass. 

In a spherically symmetric plasma with radial coordinate $R$, as we will consider in the following, resonant conversion can then occur at the radial location $R_c$ where the effective photon mass equals the DM mass,
\begin{equation}
\omega_{\rm pl}(R_c)=m_\alpha ,
\end{equation}
with $\alpha = a$ (ALP) or $\alpha = A'$ (DP).  
Tiny modifications to the plasma properties induced by the magnetic field are neglected in the computation of the conversion probability, and the plasma density is assumed to vary slowly along the radial direction. In this case, when the plasma varies on scales much larger than the de Broglie wavelength of the DM particle, the  Wentzel–Kramers–Brillouin (WKB) approximation can be used and the conversion probability for a DM particle travelling radially outwards can be written as (see e.g.~\cite{An_2021})
\begin{equation}
\label{eq:Pagamma_general}
P_{\alpha \to \gamma} \;\simeq\; f_{\rm pol}\,\pi \,\frac{g_{\rm eff}^2\, m_\alpha\,H_c}{v_r}
\, ,
\end{equation}
with $v_r = \mathbf{v}_{\alpha} \cdot \hat R$ the DM radial velocity at the resonance. The effective coupling is \(g_{\rm eff} = \epsilon, \;g_{a\gamma\gamma}|\mathbf{B}_T|/m_a\), for DPs and ALPs, respectively. We further define
\begin{equation}\label{eq:Hcdef}
H_c^{-1}\equiv\left|
\frac{\partial \ln \omega_{\rm pl}^2(R)}{\partial R} \right|_{R = R_c}\,.
\end{equation}
Finally, the polarisation factor $f_{\rm pol}$, taking the  value $f_{\rm pol} = 1$ for ALPs and $f_{\rm pol} = 2/3$ for DPs, accounts for the fact that 
in the vector case only the two transverse degrees of freedom of the DP converting into photons are propagating (see e.g.~\cite{2023Univ....9..142A}).  Note that the conversion probability scales proportionally to the scale height $H_c$. This is different from the well-known scalings manifested by ALPs and DP in the coherent, non-resonant regimes, and are peculiar of the resonant conversion mechanism considered here. Here, the probability can be understood as the ratio  between the outgoing photon energy flux across an infinitesimal surface element  at the resonance layer and the incoming DM energy flux through the same element, i.e. the ratio of the moduli of the respective Poynting vectors~\cite{Gines:2024ekm,mcdonald2024axionphotonmixing3dclassical}.

\subsection{Modelling of plasma and magnetic profiles}
In the following, we adopt different conventions for the coordinate describing the radial height, depending on the available models for the different astrophysical bodies. 
For the Sun, the density and magnetic field profiles are usually expressed as a function of the distance from the centre, normalised to the solar radius $R_\odot$, so the most commonly used  variable is the radius $R$ or  the dimensionless variable $r = R / R_\odot$. Notably for planets, it is also customary to define  the altitude above the  surface $h = R - R_s$, $R_s$ being the radius of the relevant spherical surface.  We will also resort to the normalised height variable $\bar{h}=h/H_{\rm eff}$, with 
$H_{\rm eff}$ a reference altitude specified in each case.
Finally, in order to help the reader to build an intuition for the magnitude of the conversion probabilities, in the following we present analytical approximations of the logarithmic gradients of $\omega_{\rm pl}^2$ (see Eqs.~\eqref{eq:Pagamma_general},\eqref{eq:Hcdef}). Note,  however, that 
the numerical evaluation of the expected DM-induced signal relies on the full density models of each target, without resorting to  the analytical approximations.

\subparagraph{The Sun.}
The plasma frequency profile in the corona of the Sun crucially depends on the electronic density \(n_e\) (see Eq.~\eqref{eq:plasma_frequency}). We employ the electron density model provided in \cite{Wexler_2016}:

\begin{equation} \label{eq:solar density model}
    n_e(r) = \left[\,\dfrac{65}{r^{5.94}} + \dfrac{0.768}{(r - 1)^{2.25}}\,\right]\times 10^6 \:\;\rm{cm}^{-3},
\end{equation}
which is an hybrid model from two parametrisations\;\cite{2010ApJ...722.1495H,2015A&A...583A.101M} derived using electron density data in the range [1.2--30] $R_\odot$. However, we have verified that Eq.~\eqref{eq:solar density model}
agrees with the behaviour at large radii ($> 30 \, R_\odot$) of the electron density profile in~\cite{1998SoPh..183..165L}, which is valid up to 1 AU ($\sim 215$ $R_\odot$). 
In order to cover the DM mass range probed by the instruments considered in this work, up to 
\(m_\alpha \simeq 4\times 10^{-7}\,\mathrm{eV}\), we also extend the model down to \(r = 1.12\,R_\odot\). At these radii the model yields electron densities \(n_e \sim 10^8~\mathrm{cm}^{-3}\), which are consistent (within factors of a few) with measurements of the lower corona in quiet-Sun \cite{1999JGR...104.9709F}. We thus adopt Eq.~\eqref{eq:solar density model} over the interval 
\([1.12,\,215]\,R_\odot\). The second term dictates the profile behaviour, while the first term only represents a significant correction around \(r \approx 1.6\). Hence, one infers the approximate large radius relation \(\omega_{\rm pl} \propto (r -1)^{1.125}\), and the spatial derivative of the plasma frequency computed at the resonance \(R_c\) can be expressed as
\begin{equation}
    H_c
    \simeq\dfrac{R_c-R_\odot}{2.25} \,.
\end{equation}

For the ALP case, a model for the magnetic field of the Sun in the solar corona is also needed. We employ the analytical DQCS model of Ref.~\cite{Banaszkiewicz_1998}, 
which includes a quadrupole correction to the usual dipole approximation. 
For the numerical estimates below, we take as the relevant transverse 
magnetic field for ALP conversion the component in the ecliptic plane,
\begin{equation}
    \mathbf{B}_T(r) = M \left[
-\frac{1}{r^3}
+ \frac{9Q}{8r^5}
+ \frac{K}{\left(a_1^2 + r^2\right)^{3/2}}
\right]\hat{\mathbf{z}}\:\;\rm{G},
\end{equation}
where $M$, $Q$, $K$ and $a_1$ are parameters of the model, fixed in the following to $M=1.789$, $Q=1.5$, $a_1=1.538$, $K=1$. 

\begin{figure*}[t] 
  \centering
  \includegraphics[width=\columnwidth]{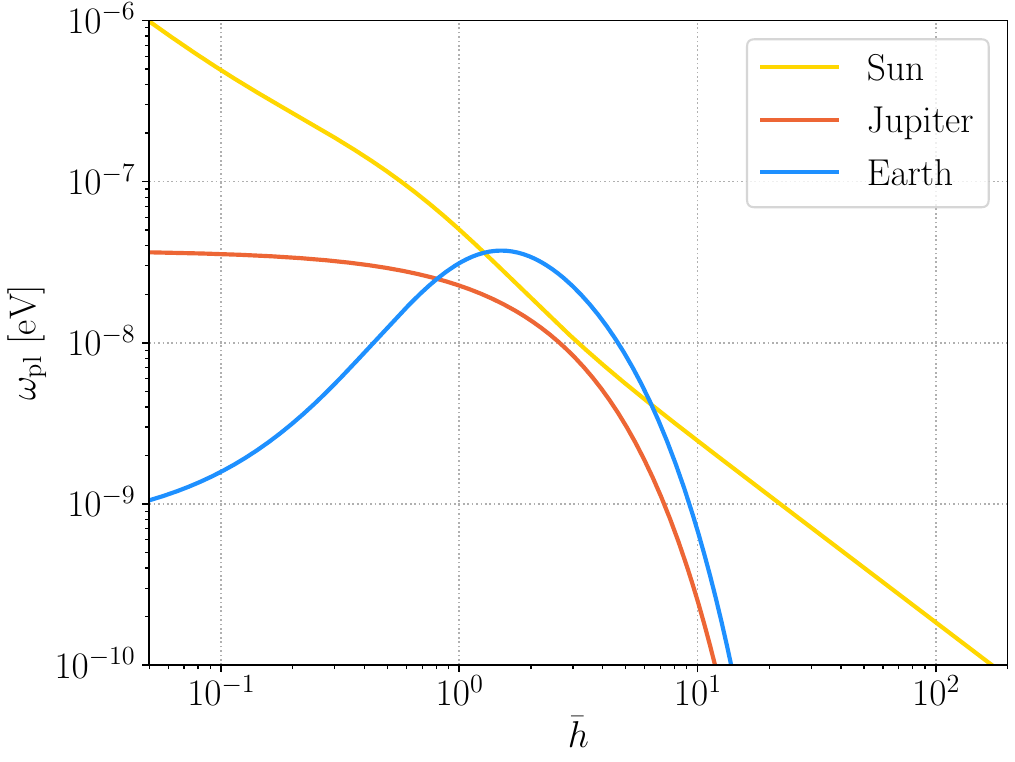}
  \includegraphics[width=\columnwidth]{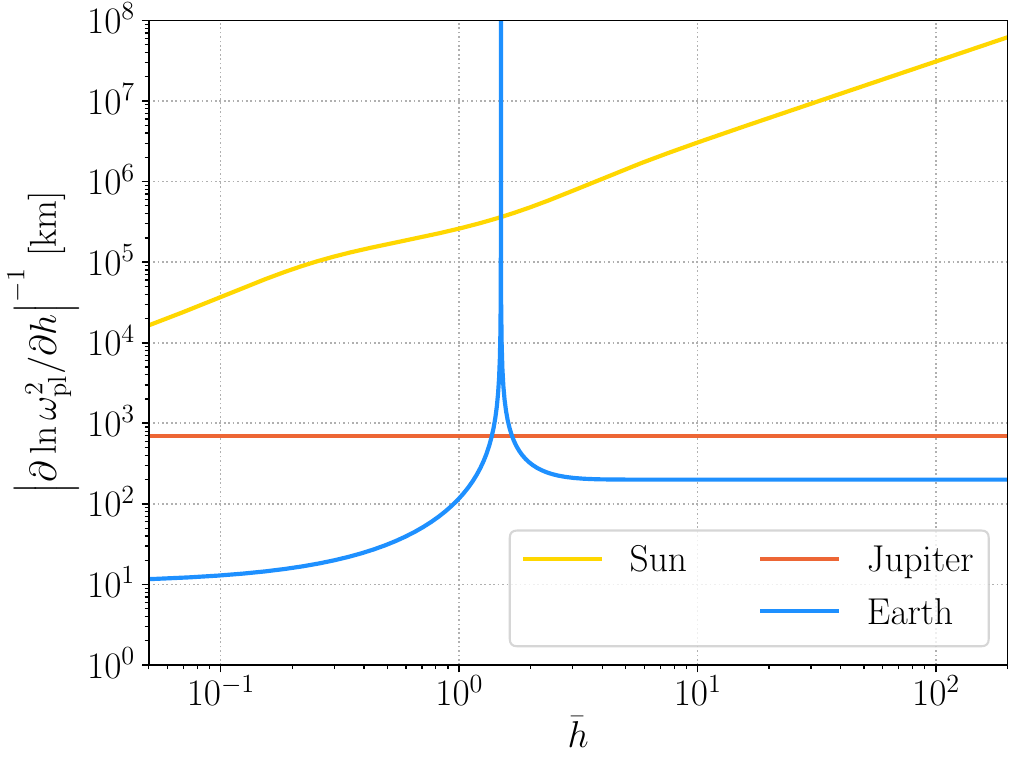}
  \caption{Plasma frequency profiles and inverse logarithmic gradients of $\omega_{\rm pl}^2$ for the Sun, Jupiter, and the Earth, expressed in terms of $\bar{h}=h/H_{\rm eff}$, where the effective heighs are $H_{\rm eff}=R_\odot, H_\oplus, H_J$ in the three cases. 
  }
  \label{fig:omega_pl_targets}
\end{figure*}

\begin{figure}[t] 
  \centering
  \includegraphics[width=\columnwidth]{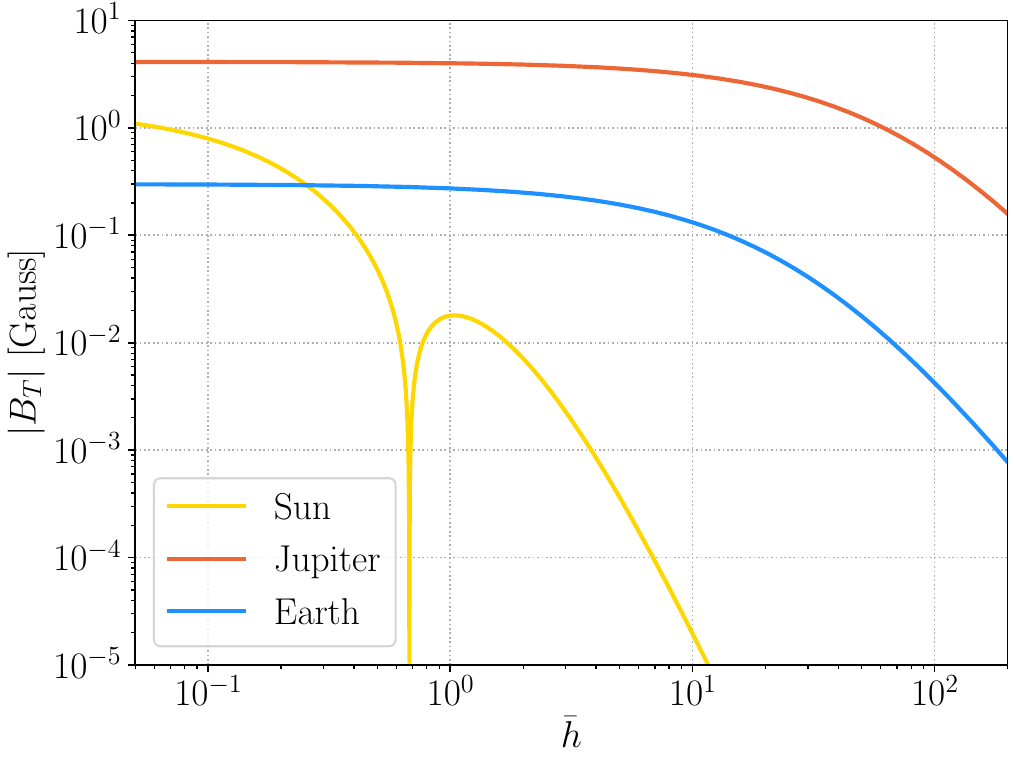}
  \caption{Transverse magnetic field profiles $|\mathbf{B}_T|$ for the Sun, Jupiter, and the Earth, in units normalised as in Fig~\ref{fig:omega_pl_targets}. In the adopted solar model, the value of the transverse field changes sign around $\bar{h}\simeq 0.7$.}
  \label{fig:Bfield_targets}
\end{figure}

\begin{figure*}[t!]
  \hspace*{-1.25cm}  
  \begin{minipage}{2.25\columnwidth}
    \centering
    \includegraphics[width=\linewidth]{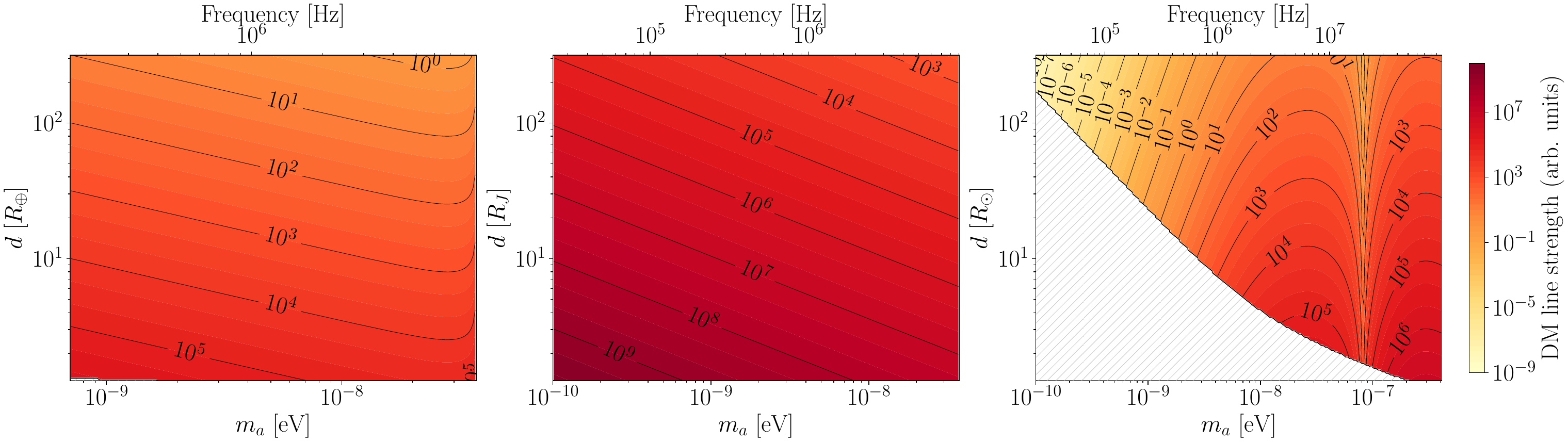}
    \caption{DM ALP line strength maps for the three targets: the Earth, Jupiter and the Sun, from left to right.}
    \label{fig:pheno_ALP}
  \end{minipage}
\end{figure*}

\begin{figure*}[t!]
  \hspace*{-1.25cm}  
  \begin{minipage}{2.25\columnwidth}
    \centering
    \includegraphics[width=\linewidth]{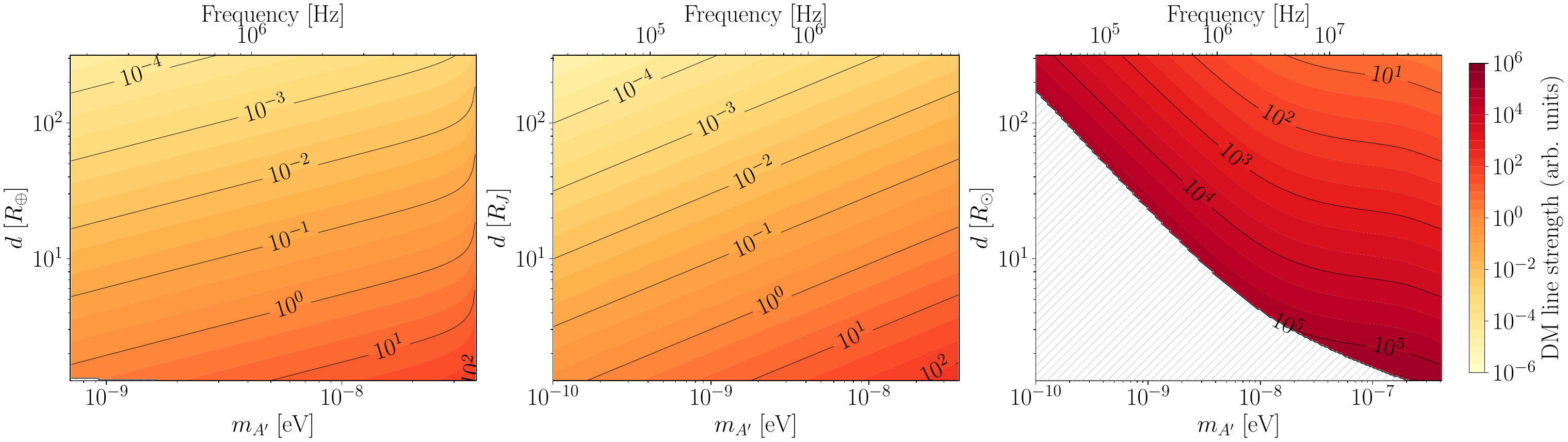}
    \caption{DM DP line strength maps for the three targets: the Earth, Jupiter and the Sun, from left to right.}
    \label{fig:pheno_DP}
  \end{minipage}
\end{figure*}

\subparagraph{Jupiter.}
In the upper ionosphere, the observed electron densities can be modeled using an exponentially decaying function with the altitude \cite{AgiwalO2025EDiJ}, 
\begin{equation}
    n_e(h) = n_0 \cdot \exp({-h/H_J}),
\end{equation}
where \(n_0\) is the electron density at the conventionally defined Jupiter radius and \(H_J\) represents a scale height related to the temperature and the masses of the ions present in the ionosphere. In \cite{AgiwalO2025EDiJ}, three values of \(H_J\) have been provided, \(H_J = \SI{250}{km},\, \SI{700}{km}, \, \SI{1700}{km}\), depending on the trajectory and on the region of the ionosphere probed. For numerical computations, we have adopted $H_J = \SI{700}{km}$ and \(n_0 = 10^6\,\si{\centi\meter^{-3}}\). Note that, for this simple exponential scale height, one has
\begin{equation}
    H_c = H_J \, .
\end{equation}

In our numerical calculations of the ALP conversion, we employ a dipole model for Jupiter's 
magnetic field. We take as the relevant transverse magnetic field
\begin{equation}
    |\mathbf{B}_T(h)| = B_{T,0} \left[ \frac{R_J}{R_J + h} \right]^3,
\end{equation}
where $B_{T,0} = 4.2\,\mathrm{G}$~\cite{https://doi.org/10.1002/2018GL077312} 
is the equatorial magnetic field at Jupiter's radius, and 
$R_J = 71{,}492\,\mathrm{km}$~\cite{Prša_2016}.
Since the plasma frequency varies on much shorter spatial scales than the magnetic field, we verified that treating $|\mathbf{B}_T|$ as constant across the relevant resonant layers is an excellent approximation.

\subparagraph{The Earth.}
We parametrise the electron density of the ionosphere using a so-called Chapman $\alpha$-function \cite{https://doi.org/10.1029/2005RS003370}:

\begin{equation}
    n_e(h) = n_{\rm max}  e^{\frac12\left( 1 - \frac{h-h_{\rm max}}{H_\oplus} - \exp\left(-\frac{h-h_{\rm max}}{H_\oplus}\right) \right)},    
\end{equation}
where \(n_{\rm max}\) is the peak density at altitude \(h_{\rm max}\). We use the physical values \(n_{\rm max} =10^6\,\si{\centi\meter^{-3}}\), \(H_\oplus = 100\,\si{\kilo\meter}\) and \(h_{\rm max}= 300\,\si{\kilo\meter}\). As before, we are interested in the natural logarithm of the profile
\begin{equation}
H_c
=\frac{2H_\oplus}{\big|e^{-(h_c-h_{\rm max})/H_\oplus}-1\big|},
\end{equation}
which, for the descending branch (\(h_c\gtrsim h_{\rm max}+H_\oplus\)) approximately reduces to  $H_c \simeq 2H_\oplus$. 

Analogously to the Jupiter case, we adopt a dipole model for the Earth's magnetic field. The equatorial magnetic field strength at the surface ($R_\oplus = 6,371\,\mathrm{km}$)~\cite{Prša_2016} is taken to be ${B}_{T,0}=0.3\,\mathrm{G}$. As in the case of Jupiter, the magnetic field changes slowly over the conversion region, so treating $|\mathbf{B}_T|$ as constant represents an excellent approximation. 

\bigskip 

Fig.\,\ref{fig:omega_pl_targets} shows the plasma frequency profiles and the corresponding inverse logarithmic gradients of $\omega_{\rm pl}^2$ for the Sun, Jupiter, and the Earth,  expressed in units of the scale heights $H_{\rm eff}=R_\odot, H_\oplus, H_J$, respectively. 

Analogously, Fig.\,\ref{fig:Bfield_targets} shows the transverse magnetic field profiles, $|\mathbf{B}_T|$, for the Sun, Jupiter, and the Earth, expressed in units rescaled as above. Note that for the Sun, the chosen magnetic-field model features a sign reversal at $\bar{h}\simeq 0.7$.

\subsubsection{DM mass ranges}
We summarise here how the relevant DM mass ranges were chosen for each of the three targets, starting from the planetary cases.

In our models, both the Earth and Jupiter are characterised by a maximum electron density of $n_e = 10^6~\mathrm{cm^{-3}}$, corresponding to a maximum plasma frequency value of $\omega_{\rm pl}^{\rm max} \simeq 3.7\times10^{-8}~\mathrm{eV}$.
This sets an upper limit on the mass of the DM particle that can resonantly convert into photons.
The lower limit, on the other hand, is determined by the validity of the WKB approximation, which requires that at the conversion layer the spatial variation of the plasma frequency occurs on scales much larger than the de Broglie wavelength of the DM particle, i.e.
\begin{equation}
    \left|\dfrac{\partial_R \omega_{\rm pl}(R)}{\omega_{\rm pl}(R)}\right|_{R=R_c}=\dfrac{1}{2\,H_c}\ll \omega_{\rm pl}(R_c)\,v_\alpha\,,
\end{equation}
where $v_\alpha$ is the DM velocity at the conversion layer. By adopting $v_\alpha \sim 10^{-3}$, we have verified that the radial point where this condition breaks down—i.e.\ where the equality holds—corresponds to a minimum plasma frequency of $\omega_{\rm pl}^{\oplus} \simeq 7\times10^{-10}~\mathrm{eV}$ for the Earth and $\omega_{\rm pl}^{J} \simeq 1\times10^{-10}~\mathrm{eV}$ for Jupiter.

For the Sun, the WKB validity condition above holds throughout the region
\([1.12,\,215]\,R_\odot\), where we adopt the density model in Eq.~\eqref{eq:solar density model}. 
Hence, the mass range considered for the solar case, $m_\alpha \in [10^{-10},\,4\times10^{-7}]~\mathrm{eV}$, was chosen simply on the basis of the sensitivity ranges of the instruments considered in our analysis, as described in the following section. We summarise the DM mass ranges adopted for the three targets in Tab. \ref{tab:mass_ranges}.

\begin{table}[t]
    \centering
    \small
    \caption{Summary of the DM mass ranges considered for each target.}
    \label{tab:mass_ranges}
    \renewcommand{\arraystretch}{1.1}
    \setlength{\tabcolsep}{8pt}
    \begin{tabular}{lcc}
        \hline\hline
        Target & $m_\alpha^{\min}$ [eV] & $m_\alpha^{\max}$ [eV] \\
        \hline
        Earth   & $7\times10^{-10}$      & $3.7\times10^{-8}$ \\
        Jupiter & $1\times10^{-10}$      & $3.7\times10^{-8}$ \\
        Sun     & $1\times10^{-10}$      & $4\times10^{-7}$   \\
        \hline
    \end{tabular}
\end{table}

\section{Dark matter signals \& detectability}\label{sec:detection}
\subsection{Dark matter-induced flux density}\label{DMflux}
In terms of radiometric quantities, the conservation of the phase space density of photons translates in the conservation of the \textit{specific intensity} $B_\nu$ (also indicated as $I_\nu$ in the literature) along the rays in free space~\footnote{For a proof, see e.g.~Sec.~1.3 in~\cite{1986rpa..book.....R}. A generalisation to a medium with variable refraction index is stated in Problem 8.1 of Ref.~\cite{1986rpa..book.....R}, and a further generalisation accounting for the energy redshift/blueshift effect is discussed in App.~B of Ref.~\cite{Battye:2021xvt}, where one also finds the link of the quantity $B_\nu$ with the Poynting vector ${\bf S}$: $B_\nu=|{\rm  d} {\bf S}/{\rm  d}\Omega {\rm  d}\nu|$.}.
This quantity represents the power per unit area, per unit frequency, and per unit solid angle crossing a surface normal to the propagation direction.
However, a quantity of closer interest for the observations of a detector at position ${\bf x}$ pointing at direction $\hat{n}$ is the \textit{spectral flux density}, defined as
\begin{equation}
F_\nu({\bf x})=\int_{\Omega_{\rm s}} B_\nu({\bf x},\hat k) \hat{k}\cdot\hat{n}\, {\rm d}\Omega_{\hat k}\,,
\end{equation}
where the integral extends over the solid angle subtended by the source, $\Omega_{\rm s}$, and $\hat{k}$ denotes the direction of the  ray. 
Although SI units for $F_\nu$ are W/m$^{2}$/Hz,  following the convention used in radio astronomy, we will express spectral flux densities in (multiples of) Jansky (Jy), where 1 Jy = 10$^{-26}$  W/m$^{2}$/Hz.

Using the constancy of $B_\nu$ along rays in free space, we can evaluate it at the emission point ${\bf x}_e$ on the visible surface and parametrise each ray by its direction at the detector, specified by the two angles $(\theta,\phi)$; $\cos\theta=\hat k\cdot\hat n$ describes the angular distance from the centre of the observed target, and $\phi$ is the azimuthal angle around it. The spectral flux density is then
\begin{equation}
F_\nu=\int_{\Omega_{\rm s}}\! B_\nu(\theta,\phi)\,\cos\theta\,{\rm d}\Omega\,.
\end{equation}
For a surface of uniform brightness one has $B_\nu(\theta,\phi)=B_\nu^{\rm surf}=const.$ and the integral gives
\begin{equation}
F_\nu = B_\nu^{\rm surf}\int_{\Omega_{\rm s}}\!\cos\theta\,{\rm d}\Omega\,
=\pi B_\nu^{\rm surf}\left(\frac{R_c}{d}\right)^2 \,,
\end{equation}
where $R_c$ is the radius of the emitting surface (in our case, the conversion layer) and $d$ is the distance between the source centre and the observer.

In the case of DM resonant conversion, the signal brightness (directly related to the local Poynting flux of the outgoing electromagnetic wave) can be expressed as the energy flux of the incoming DM per unit solid angle and frequency, multiplied by the conversion probability (see e.g.~Eq.~(8) in Ref.~\cite{Battye:2021xvt}) as: 
\begin{equation}
B_\nu^{\rm DM}
    = 2 \ \delta(\nu-\nu_\alpha)\,
      \frac{\rho_{\rm DM}\,v_\alpha}{4\pi}\,
      P_{\alpha\to\gamma} \, ,
      \label{eq:brightness}
\end{equation}
where $\nu_\alpha$ is the photon frequency corresponding to the DM rest mass and $\rho_{\rm DM} = 0.3 \rm \; GeV/cm^3$ is the DM density at the conversion surface, which we assume to equal the local DM density. The small intrinsic width of the signal line $\Delta\nu_{\rm line}/\nu_\alpha \sim 10^{-6}$, set by the DM velocity dispersion, is neglected here. If accounting for it, an average over the velocity distribution of DM would be required. The factor of two in Eq.~\eqref{eq:brightness} arises because photons converted from both outward moving and infalling DM can be detected: the latter are reflected by the plasma layer of higher plasma frequency, which acts as a mirror~\cite{PhysRevLett.121.241102,An_2021}.

Using central radial trajectories as a proxy for the whole emitting disc,  $B_\nu^{\rm surf}=B_\nu^{\rm DM}(\theta = 0)$, $v_\alpha = v_r$, and the 
DM-induced flux density can be written in a \textit{velocity-independent} form: 
\begin{equation} \label{eq:DM flux density}
F_{\nu}^{\rm DM} \simeq \frac{\pi\, f_{\rm pol}}{2}\delta(\nu-\nu_\alpha)\,
     \rho_{\rm DM}\,
      \,g_{\rm eff}^2\, m_\alpha\,H_c\,\left(\frac{R_c}{d}\right)^2 .
\end{equation}

This is the approximation of the DM spectral flux density signal that we will use from now on in the text and for numerical estimates.

To connect with the observable signals as measured by an antenna, we 
need to introduce detector-dependent features. First, the flux density \textit{observed} by an antenna of normalised power pattern $U(\theta,\phi)$ is commonly defined 
as \cite{2016era..book.....C,1966raas.book.....K}
\begin{equation}
\label{eq:observed flux density}
F_{\nu,{\rm obs}} \;=\; \int_{\Omega_{\rm s}} B_\nu(\theta,\phi)\,U(\theta,\phi)\,{\rm d}\Omega\,,
\end{equation}
where the integration is performed over the solid angle $\Omega_{\rm s}$ subtended by the source. The function $U(\theta,\phi)$ describes the directional sensitivity of the antenna and is normalised such that its maximum value is unity. It is also useful to define the antenna beam solid angle $\Omega_{\rm A}$ as the integral of the normalised power pattern over a sphere 
\begin{equation}
     \Omega_{\rm A}=\int_{4\pi}U(\theta,\phi) d\Omega\,.
\end{equation}
For a short dipole antenna, this quantity equals to $\Omega_{\rm A}^{\rm dip}=8\pi/3$.

For compact sources ($\Omega_{\rm s}\!\ll\!\Omega_{\rm A}$) observed on boresight, i.e.\ in the direction of maximum gain, 
both $U(\theta,\phi)$ and $\cos\theta$ are approximately unity across the source solid angle and the observed flux density
coincides with the true flux density,
\begin{equation}
F_{\nu,\mathrm{obs}} \;\simeq\; F_\nu.
\end{equation}
Notice that this equality breaks down for other environmental emissions, considered later in this work.

Because the DM signal is monochromatic, in Figs.~\ref{fig:pheno_ALP}–\ref{fig:pheno_DP} we show maps of the DM \emph{line strength}—the coefficient of the Dirac delta in Eq.~\eqref{eq:DM flux density}—in the $(m_\alpha,\,d)$ plane, expressed in an  arbitrary unit which differs between the two models, but is common among the three targets.
These plots highlight the regions where the physical signal is largest while remaining independent of instrument or observation–specific features (e.g.\ channel resolution $\Delta\nu$ and integration time $t_{\rm int}$); these effects will be taken into account in the next subsection when assessing detectability for given instruments.
In Fig.~\ref{fig:pheno_ALP}, we display the DM line strength for the ALP case, while in Fig.~\ref{fig:pheno_DP} the one
for the DP case, for the three targets of interest: the Earth, Jupiter and the Sun, from left to right.
For each target, the mass range is limited from below by requiring the 
validity of the WKB approximation and from above by the maximum of the plasma frequency, as we explained in Sec.~\ref{sec:signals}.
For all three targets, the DM line strength scales with \(d^{-2}\), as expected from the geometric dilution factor.

For both the Earth and Jupiter, the quantities $H_c$ and $R_c$ remain approximately constant across the mass ranges of interest. This follows from the fact that the plasma frequency varies on spatial scales much shorter than the planetary radius. Since the plasma frequency scale is also much shorter than the one over which the magnetic field varies, in the ALP case $|\mathbf{B}_T|$ can be treated as nearly constant with mass. Consequently, the DM line strength differences among the two targets arise primarily because of the different values of $|\mathbf{B}_T|$, $H_c$ and $R_c$ at the conversion layer.
Quantitatively, the Jovian magnetic field exceeds the Earth one by more than an order of magnitude, and its length scale $H_c$ is almost four times larger. Since the ALP line strength scales roughly as $|\mathbf{B}_T|^2 H_c$, this leads naturally to a signal roughly three orders of magnitude stronger for Jupiter than for the Earth.
For DPs, the situation differs because the effective coupling has a different structure, $g_{\rm eff} = \epsilon$ instead of $g_{\rm eff} = g_{a\gamma\gamma}|\mathbf{B}_T|/m_\alpha$ for ALPs. This distinction not only modifies the mass scaling of the signal but also removes the magnetic-field dependence. Consequently, the contrast between the Jupiter and Earth cases is milder for DPs.

For the Sun (rightmost panel of Figs.~\ref{fig:pheno_ALP} and~\ref{fig:pheno_DP}), the situation is more complex as both $H_c$ and $R_c$ scale with the DM mass. For both DM models, decreasing $m_\alpha$ causes the resonance condition $\omega_{\rm pl}(R_c)=m_\alpha$ to move the conversion layer progressively farther away from the Sun, increasing $R_c$. As a direct consequence, for each mass there exists a minimum observable distance $d_{\min}=R_c(m_\alpha)$, resulting in an effectively inaccessible region in the $(m_\alpha,d)$ parameter space, indicated by the hatched grey region in the figures. Notice that in the planetary case, this effect is not visible because of the short scale of the plasma frequency variation.
Moreover, both the geometrical scaling factor $(R_c / d)^2$  and $H_c$ grow at small masses. For DP, this scaling combines with the linear dependence on $m_{A'}$, producing the line strength map seen in the rightmost map of Fig.~\ref{fig:pheno_DP}. Overall, across the entire parameter space, the DP line strength appears stronger than in planetary scenarios owing to the higher value of $H_c$. In the ALP case, however, the quadratic dependence on the magnetic field dominates the structure of the strength map. It results in a sharp suppression of the signal at small masses and produces an apparent null sensitivity around $m_a \sim 10^{-7} \, \mathrm{eV}$. This feature originates from a singularity in the magnetic-field model used, and should be interpreted as an artefact of the input profile merely indicating a reduced sensitivity.


\subsection{Radio sensitivity}
Since, in the frequency range under consideration, radio receivers measure the power integrated over a finite spectral channel $\Delta\nu\gg \Delta\nu_{\rm line}$, it is convenient to define the \textit{channel-averaged flux density}:
\begin{equation}\label{eq:average_SFD}
    \begin{aligned}
            \bar{F}^{\rm DM}(m_\alpha,& g_{\rm eff};\, d) =\frac{1}{\Delta \nu} \int_{\Delta\nu} F_{\nu}^{\rm DM}\,{\rm d}\nu \\
            \simeq& \frac{\pi\, f_{\rm pol}}{2\, \Delta \nu}\,
     \rho_{\rm DM}
      g_{\rm eff}^2\, m_\alpha\,H_c(m_\alpha)\,\left(\frac{R_c(m_\alpha)}{d}\right)^2 .
    \end{aligned}
\end{equation}

This is the relevant quantity to compare with the noise fluctuations detected by the receiver. 
A convenient measure of the latter is the \textit{system equivalent flux density} (SEFD)\footnote{In radio astronomy, the SEFD is typically expressed as 
$\mathrm{SEFD} = {2 k_\mathrm{B} T_\mathrm{sys}/A_\mathrm{eff}}$, 
where $T_\mathrm{sys}=T_{\rm rx}+T_{\rm env}$ is the system temperature, composed by the receiver noise and every contribution from the environment. $A_\mathrm{eff}$ is the effective collecting area of the antenna or the array.}.
It can be operatively defined as the flux density of a hypothetical radio source that would produce, at the output of the receiving system, a power equal to the system’s own thermal noise.
With this definition, the detectability condition of the DM signal for a required signal-to-noise ratio SNR becomes
\begin{equation}\label{eq:radiometer}
\bar{F}^{\rm DM}(m_\alpha, g_{\rm eff};\, d) \;\ge\;
\frac{\rm SNR}{\sqrt{\Delta\nu \, t_{\rm int}}}\,
{\rm SEFD}\,,
\end{equation}
where $\Delta\nu$ is the observed bandwidth of the measurement, and $t_{\rm int}$ is the integration time. For the case of an interferometric system (i.e.~array of antennas), the SEFD scales also as the inverse of $\sqrt{N_{\rm ant}  (N_{\rm ant}- 1)}$, with $N_{\rm ant}$ as the number of antennas \cite{2016era..book.....C}.
In general, the SEFD may encode multiple sources of noise --including receiver noise, atmospheric absorption and sky backgrounds. 

Since we restrict ourselves to the case of space-based antennas, we consider two main contributions:
\begin{equation} \label{eq:SEFD}
{\rm SEFD} \;=\; {\rm SEFD_{rx}} \;+\; {\rm SEFD_{env}},
\end{equation}
where ${\rm SEFD_{rx}}$ represents the intrinsic receiver noise, while ${\rm SEFD_{env}}$ includes the environmental contributions discussed below.
Additionally, if the source has a strong radio continuum (and stationary) emission, there should be also a term
associated to the source emission. With the scope of illustrating the ultimate reach of current and future instruments, 
we will neglect this contribution in what follows.

We will compute sensitivity curves in the $(m_\alpha, g_{\rm eff})$ parameter space by solving the implicit equation in Eq.~\eqref{eq:radiometer}, for fixed observation distances $d$ and instrument specifications corresponding to the receivers listed below. In all sensitivity calculations, we fix SNR = 2. 

\subsubsection{Environmental noise}
The main astrophysical contribution to the environmental noise originates from diffuse synchrotron emission of cosmic–ray electrons in the Galactic magnetic field, which are partially absorbed by free–free processes in the warm ionised medium below  $\sim$10 MHz~\cite{1978ApJ...221..114N,Peterson_2002}. 
Regarding this background, we adopt the empirical spectrum derived by Novaco and Brown~\cite{1978ApJ...221..114N}, 
based on measurements of the non-thermal radio emission with the RAE-2 spacecraft.
It can be parametrised as
\begin{equation}
B_\nu^{\rm gal} \;=\; B_0 \,\nu^{-0.76}\, \exp\!\left(-\,3.28\, \nu^{-0.64}\right),
\end{equation}
where the frequency $\nu$ is expressed in MHz and the normalisation constant is 
$B_0 = 1.38\times10^{-19}\;\mathrm{W/m^{2}/Hz/sr}$. 

A second, local component to the noise arises instead from the plasma surrounding the antenna. In a thermal medium at temperature $T$, one consequence of the fluctuation–dissipation theorem is that a circuit with resistance $R$, even in absence of an applied current, manifests a stochastic voltage, whose variance ${\cal V}^2$, measured over a frequency bandwidth $\Delta \nu$, is ${\cal V}^2=4\,k_B\, T\,R\,\Delta \nu$; this is the so-called \textit{Johnson–Nyquist noise}~\cite{1928PhRv...32...97J,1928PhRv...32..110N}. 
Similarly, an antenna in a plasma experiences fluctuations in its potential, whose variance depends on the temperature, the plasma density, as well as the geometry of the antenna. This effect has been used for the local plasma diagnostics in the study of different environments in the Solar System, and is known as quasi-thermal noise (QTN) (see e.g.~\cite{Meyer2017}). If we define $L_D$ as the plasma Debye length of the plasma and $L$ as the length of a single antenna wire,
the variance of voltage per unit frequency (also known as voltage spectral density~\cite{2016era..book.....C}) $V_{\rm QTN}^2$ can be expressed in the limit of our interest, i.e.~$2\pi\,\nu  \gg \omega_{\rm pl}\,(L_D/L)$, as (see e.g. Eq.~(46) in~\cite{Meyer2017}): 
\begin{equation}
\label{eq:V_sqrd QTN}
\begin{aligned}
    V_{\rm QTN}^2(\nu)\simeq \frac{ \omega_{\rm pl}^2k_BT}{4\pi^3  \nu^3 L}=
4\times 10^{-5}\frac{n_e\,T_e}{\nu^{3}\,L}
\quad{\rm V^2/Hz},
\end{aligned}
\end{equation}
where  $n_e$  and $T_e$  are the local electron density and temperature and $\nu$  is the considered frequency. In the second equality, $n_e$, $T_e$,  $\nu$ and $L$ are expressed in $\mathrm{cm^{-3}}$, $\mathrm{K}$, $\mathrm{Hz}$, and $\mathrm{m}$, respectively. For the local plasma density and temperature, we will use representative values of $n_e$ and $T_e$ corresponding to $1\,{\rm AU}$, for observations made in orbit close to the Earth (or the Moon) and $5\,{\rm AU}$ for measurements close to Jupiter, as reported in Ref.~\cite{2024ApJ...963...46L}. In the case of observations made at variable distances from the Sun,
we use the density model $n_e(r)$ in Eq.~\eqref{eq:solar density model}
to account for the distance dependence of the QTN in Eq.~\eqref{eq:V_sqrd QTN}. For the electron temperature, we verified that adopting the profile $T_e(r)$ in\;\cite{2009LRSP....6....3C} leads to negligible corrections for distances above 2 $R_\odot$.

As a result of these two contributions, the environmental voltage spectral density at the terminals of a short antenna writes as~\cite{2011RaSc...46.2008Z} 
\begin{equation}
    V_{\rm env}^2=V_{\rm QTN}^2 + \frac12 \Omega^{\rm dip}_{\rm A} \, Z\,l_{\rm eff}^2B_{\nu}^{\rm gal}, \label{eq:Va}
\end{equation}
where $Z $ is the medium impedance, which in what follows we assume equal to the vacuum value $Z_0 = 1$ in natural units, or $377\,\Omega$ in SI units. Moreover, we define with $l_{\rm eff}$ the \textit{effective electric length of an antenna}, which is given by the ratio of the voltage induced at the terminals and the electric field of an incoming electromagnetic plane wave whose field direction is parallel to the antenna. 

The environmental SEFD, or SEFD$_{\rm env}$, then simply reads \cite{2011RaSc...46.2008Z,2021A&A...656A..33V}:
\begin{equation} \label{eq:SEFD_env}
    \mathrm{SEFD_{env}} = 2\frac{V^2_{\rm env}}{ Zl_{\rm eff}^2} \, .
\end{equation}
In what follows, we will approximate $l_{\rm eff} \simeq L$.
More specifically, for the calculation of the sensitivity in the presence of environmental-noise only, cf.~Sec.~\ref{sec:results_env}, we will fix $t_{\rm int}=100~{\rm hr}$, $\Delta\nu=1~{\rm kHz}$ and $L=5~{\rm m}$.

\subsubsection{Instrumental effects}
A further instrumental complication is that what is measured at the receiver input is not the potential drop at the antenna, but 
\begin{equation}
    V_r^2=V^2_{\rm rx}+\Gamma^2 V_{\rm a}^2 \, , 
\end{equation}
where $V^2_{\rm rx}$ is the voltage noise spectral density measured at the receiver, 
\(\Gamma\) is the voltage transfer factor due to the impedance coupling in the antenna‐receiver system, 
and  \(V_{\rm a}^2 = V_{\rm env}^2 + V_{\rm DM}^2\) is the antenna voltage spectral density containing the environmental terms as well as any sky signal, including the DM one.

\begin{table*}[t]
\centering
\small
\caption{Instruments specifications used for the SEFD calculations: (1) name; (2) frequency range, broken in the frequency sub-bands when indicated; 
(3) antenna length; (4) $\Gamma l_{\rm eff}$; 
(5) spectral resolution, per frequency sub-band; (6) SEFD$_{\rm rx}$, indicating 
the individual antenna (or full array) equivalent flux density, except for the instruments marked by $^{*}$, 
which already include environmental background contributions (see main text for details). We stress that some of the 
SEFD reported in this table are approximated values which suffer from systematics in the extraction of the data points from the 
corresponding references.
}
\label{tab:instrument_specs}
\setlength{\tabcolsep}{5pt}
\renewcommand{\arraystretch}{1.12}
\begin{tabular}{lccccc}
\hline\hline
Instrument & $\nu$~range [MHz] & $L$  [m] & $\Gamma l_{\rm eff}$ [m] &Spectral res. [kHz] & SEFD$_{\rm rx}$ [Jy]\\
\hline
\multicolumn{5}{l}{\textbf{Lunar probes}}\\
NCLE & 0.1--60 & $5$& -- &  1 & $7\times 10^{4} - 1.0\times 10^9$ \\
DEX$^{*}$  & 1--100   & -- & -- & 12  & $7\times10^{2} - 3\times10^{4}$ \\ 
FARSIDE$^{*}$ & 0.1--40 & -- & -- & 28.5  & 0.23(28)$\times10^{3}$ at 200~kHz (15~MHz) \\ 
FarView$^{*}$ & 5--50 & -- & -- & 5  & 42(7.0) at 15~kHz (40~MHz) \\
[2pt]
\multicolumn{5}{l}{\textbf{Solar probes}}\\
SO RPW & 0.004--1.0, 0.4--16.4 & $6.5$& 3.24 & 18--46, 50--100  & $ 2 \times 10^7 - 9 \times 10^8$ \\
PSP FIELDS & 0.01--19 & $2$ & 1.17 & 45--855 & $ 3 \times 10^6 - 4 \times 10^7$ \\ 
[2pt]
\multicolumn{5}{l}{\textbf{Planetary probes}}\\
Cassini RPWS & 0.003--0.3, 0.125--16 & $10$& -- & 0.15--15, 3--6  & $ 6 \times 10^5 - 1.2 \times 10^8 $ \\
Juno/WAVES & 0.01--3, 3--41 & $2.4$ & -- & 1--410, 1000 & $2\times10^7-7\times10^{12} $ \\ 
JUICE/RWI & 0.08--45 & $3$& -- & 2--1127& $1.0\times 10^{8} - 6 \times 10^{8}$ \\ 
\hline
\end{tabular}
\end{table*}

The receiver noise is an intrinsic property of each instrument. In some cases it is quoted in the literature directly as equivalent flux density (in Jy or $\mathrm{W/m^2/Hz}$), in others it is given as electric-field spectral density $E_{\rm rx}^2$ (in $\mathrm{V^2/m^2/Hz}$) calibrated at the antenna terminals, or it is provided as voltage noise spectral density measured at the receiver,
\(V_{\rm rx}^2\).
Note that, by definition, the electric field density and the voltage density are related by $E_{\rm rx}^2 = V^2_{\rm rx}/(l_{\rm eff}\Gamma)^2$.

Analogously to Eq.~\eqref{eq:SEFD_env}, to translate the receiver contribution to an equivalent flux at the antenna we use the equation:
\begin{equation} \label{eq:SEFDrx}
    \,{\rm SEFD_{rx}} \;=\; \frac{2\,V_{\rm rx}^2}{Z\big(\Gamma\,l_{\rm eff}\big)^2}\, = 2 \dfrac{E_{\rm rx}^2}{Z
    }\, .
\end{equation}

The following paragraphs summarise the main properties of past, current and planned detectors considered in our analysis—namely, their frequency coverage, frequency resolution, and nominal receiver sensitivity. For clarity, we group them into three categories: lunar, solar, and planetary probes, all operating in the low–frequency radio domain.
In Tab.~\ref{tab:instrument_specs}, we report the instruments specifications and the computed SEFD$_{\rm rx}$, when not directly available 
in the literature.

For instruments with overlapping sub–bands, we adopt, for the sensitivity estimates, 
the configuration with the finest available spectral resolution (i.e.\ the smallest 
channel bandwidth) within the relevant frequency range.

\medskip

\textbf{Lunar probes.} The Netherlands–China Low–Frequency Explorer (NCLE)~\cite{Karapakula_2024}, onboard the Chang’e–4 mission at the Earth–Moon L2 point, consists of three monopole antennas, each 5~m long. It operates from 80~kHz to 80~MHz with a spectral resolution of 1~kHz. Its analog receiver sensitivity, i.e.~SEFD$_{\rm rx}$, is provided in Figs.~21--23 of Ref.~\cite{Karapakula_2024} in ${\rm W/m^{2}/Hz}$. 
It does not include the Galactic nor the QTN noise. 

The Dark Ages Explorer (DEX)~\cite{2025arXiv250403418B}, a planned lunar–farside array ($N_{\rm ant}$ = 1024), will cover 1--100~MHz, with a 12~kHz spectral resolution. Fig.~4 of Ref.~\cite{2025arXiv250403418B} reports the sensitivity in ${\rm W/m^{2}/Hz}$ for $t_{\rm int}$=1~hr and a frequency resolution $\Delta\nu = 0.5\nu$ (derived from Fig.~4). It already includes the sky–noise contribution from the Galactic radio background, the QTN being negligible at 
these frequencies. We compute
the full array SEFD by rescaling for the observing time and bandwidth.

The Farside Array for Radio Science Investigations of the Dark Ages and Exoplanets (FARSIDE)~\cite{2021PSJ.....2...44B} is a planned array of 128 dipole pairs to be deployed on the lunar surface. It will operate in the 0.1--40~MHz range, with 28.5~kHz spectral resolution. The full array sensitivity is expected to be 230 Jy at 200 kHz and 2.8 $\times 10^4$ Jy at 15 MHz (Tab.~1 of~\cite{2021PSJ.....2...44B}). QTN noise and Galactic background are included in the sensitivity calculation.

FarView~\cite{2024AdSpR..74..528P} is a concept for an in-situ manufactured lunar farside array comprising $\sim10^5$ dipoles operating in the 5--50~MHz band, with a nominal spectral resolution of 5~kHz. The design goal is to reach a sensitivity of $\sim$2~mJy at 15~MHz and $\sim$0.2~mJy at 40~MHz for 1 minute of integration with $\Delta \nu $=0.5$\nu$, with $\nu$ as the centre frequency of the band. The Galactic sky noise is included in this estimate. We compute
the full array SEFD by rescaling for the observing time and bandwidth.

We consider NCLE to set sensitivity projections for the three targets under study, 
while lunar arrays are used for Jupiter and the Sun only, as they are planned to be placed on the far side of the Moon.

\medskip
\textbf{Solar probes.}
The Solar Orbiter (SO) RPW instrument~\cite{2020A&A...642A..12M} operates from 10~kHz to 40~MHz using three 6.5~m monopole antennas. The spectral resolution is 4.5\% in the 4~kHz--1.024~MHz range, 50~kHz for 0.4--3.6~MHz, and 100~kHz for 3.6--16.4~MHz.  
The reported sensitivities are given in ${\rm V/Hz^{\frac12}}$ (Fig.~27 of Ref.~\cite{2020A&A...642A..12M}). These values describe $V_{\rm rx}$, and do not include Galactic or QTN components. Using Eq.~\eqref{eq:SEFDrx}, we compute the 
SEFD$_{\rm rx}$, given the value of $\Gamma l_{\rm eff}$.
The SO reaches perihelion distances down to $d\sim$0.28~AU ($\sim$60~$R_\odot$) from the Sun. We use these configurations to estimate its sensitivity to DM signals from the Sun.

The FIELDS instrument on Parker Solar Probe (PSP)~\cite{2016SSRv..204...49B,2016JGRA..121.5088M} employs two 2~m dipoles and covers the range 10~kHz--19~MHz, with a frequency fractional resolution of 4.5\%.  
The measured receiver noise $V_{\rm rx}$ is shown in Fig.~3 of Ref.~\cite{2017JGRA..122.2836P}. As for SO case, we convert this into SEFD$_{\rm rx}$.
The PSP spacecraft performs repeated perihelion passes reaching distances of $d\sim$0.046~AU ($\lesssim$10~$R_\odot$) from the Sun. We use these configurations to estimate its sensitivity to DM signals from the Sun. 

We highlight here that both solar probes have sensitivities dominated by instrument noise.

\medskip
\textbf{Planetary probes.}
Cassini RPWS~\cite{2004SSRv..114..395G} operated from 1~Hz to 16~MHz using three nearly orthogonal monopoles, each 10~m long, and could achieve fractional frequency resolutions of 5\% in the low-frequency range (3.5--319 kHz) and between $\sim$3--6 kHz in the high-frequency band (125 kHz--16.125 MHz).
The in–flight noise level $E^2_{\rm rx}$ is given in Fig.~24 of Ref.~\cite{2004SSRv..114..395G}, and already includes the environmental contributions.
Cassini performed flybys with its radio instruments operating above Earth ($d\sim\, 1.2\,R_{\oplus}$, 18 August 1999) and Jupiter ($d\sim\, 137\,R_{J}$, 30 December 2000). We therefore use Cassini to derive sensitivity curves for DM signals from both the Earth and Jupiter.

Juno/WAVES, launched in 2011, operates from 10~kHz to 41~MHz with a monopole length of $L\simeq2.4$~m (half the 4.8~m tip–to–tip separation). The receivers in the low-mid frequency range (50 Hz--3 MHz) achieve a fractional frequency resolution of 1.37\%, and a linear frequency resolution of 1 MHz in the high frequency range (3 MHz--41 MHz).
Its in–flight noise $E^2_{\rm rx}$ is reported in Fig.~27 of Ref.~\cite{2017SSRv..213..347K}, and includes environmental contributions. 
The Juno spacecraft performed an Earth flyby at $d\,\sim\,1.1\,R_{\oplus}$ on 9 October 2013 and is currently operating in orbit around Jupiter, with perijoves at distances $d \lesssim 1.1\,R_J$. We therefore use Juno to derive sensitivity curves for DM signals from the Earth and Jupiter.

Finally, the Jupiter Icy Moons Explorer (JUICE)~\cite{2025SSRv..221....1W} was launched in 2023
and it is planned to enter a Jovian orbit in Summer 2031.
The instrument operates from 80~kHz to 45~MHz using three 3~m monopoles, with a frequency fractional resolution of 2.5\%. We adopt the expected electric-field noise, $E_{\rm rx}$, in Fig.~17 of Ref.~\cite{2025SSRv..221....1W}. The JUICE spacecraft performed an Earth flyby at $d\sim 2\,R_{\oplus}$ on 20 August 2024 and it is expected to perform 67 perijove passes at distances between $\sim$9 and 20~$R_J$. We therefore use JUICE to derive sensitivity curves for DM signals from both targets.

\bigskip 

To quantify final sensitivities, we adopt the full $\mathrm{SEFD}$ as in Eq.\;(\ref{eq:SEFD}), and the nominal frequency resolution $\Delta\nu$ for each instrument in the table.

In the case of lunar probes and arrays, the distance $d$ from the target --and thus the corresponding DM flux density $\bar{F}^{\rm DM}$-- can be considered constant over typical integration times of $\sim 100$~hr. Therefore, for these instruments we solve the radiometer equation in Eq.~\eqref{eq:radiometer} by setting $t_{\rm int}=100$~hr and evaluating $\bar{F}^{\rm DM}$ to a specified value of $d$~\footnote{Specifically, for all the lunar instruments we adopt $d_\odot = 1~{\rm AU}$ for the Sun and $d_J = 4~{\rm AU}$ for Jupiter. The NCLE, which operates near the Earth--Moon L2 point, can also observe the Earth; for this case we adopt $d_\oplus = 4.5\times10^{5}~{\rm km}$.}.

However, using a constant value of DM flux density over hundreds of hours may be inaccurate for solar and planetary probes, for which the distance to the target can vary significantly over short timescales. This is particularly relevant for the spacecrafts performing only a single flyby near the target (e.g.\ Cassini at Jupiter and Earth, Juno and JUICE at Earth). For these cases, we obtain the time-dependent trajectories $d(t)$ of the spacecraft from NASA's JPL \textit{Horizons} system \footnote{\url{https://ssd.jpl.nasa.gov/horizons/app.html}}, and derive the corresponding sensitivity curves using the general time-dependent form of the radiometer equation:
\begin{equation}
\label{eq:time_dep_radiometer}
{\rm SNR}^2 \;=\;
\frac{\Delta\nu}{\mathrm{SEFD}^2}
\int_{\bar{t}-\Delta t/2}^{\bar{t}+\Delta t/2}
\bar{F}_{\rm DM}^2(t)\,{\rm d}t\,,
\end{equation}
where $\bar{t}$ denotes the moment of closest approach. We verified that, for Earth flybys, integration times of order $\Delta t \sim1~{\rm hr}$ already saturate the sensitivity limit, beyond which longer integrations yield negligible improvement.
The Cassini flyby above Jupiter, on the other hand, lasted significantly longer, with the spacecraft remaining near its closest approach for several days ($d\sim 137$--$140\,R_J$). Consequently, we adopt an effective integration time of $\Delta t \simeq 100~\mathrm{hr}$ for this case.
For spacecraft performing multiple pericentre passages around their respective targets (SO and PSP around the Sun, JUICE and Juno around Jupiter), it is reasonable to assume a cumulative observing time of order $100$~hr near pericentre over multiple orbits. Accordingly, we adopt $t_{\rm int}=100$~hr and fix $d$ to the minimum approach distance reported previously for each of these spacecrafts.

\section{Results}\label{sec:results}
\subsection{Environmental noise}\label{sec:results_env}
As a first step, we assess how the environmental noise alone limits the sensitivity to ALP or DP DM. This serves us to evaluate the exclusion and discovery potential in presence of this irreducible
noise contribution, and to illustrate the ultimate reach under ideal instrumental performances or for sky-noise dominated single antennas. 
This will be the case for the lunar arrays considered in this work, whose sensitivity benefits from the rescaling due to the large number of antennas.

With the bandwidth and integration time choice quoted above, we solve Eq.~\eqref{eq:radiometer} considering only the environmental noise contribution. The procedure is repeated for several values of $d$, and for both ALPs and DPs across the three targets.

\begin{figure*}[t] 
  \centering
  \includegraphics[width=2.1\columnwidth]{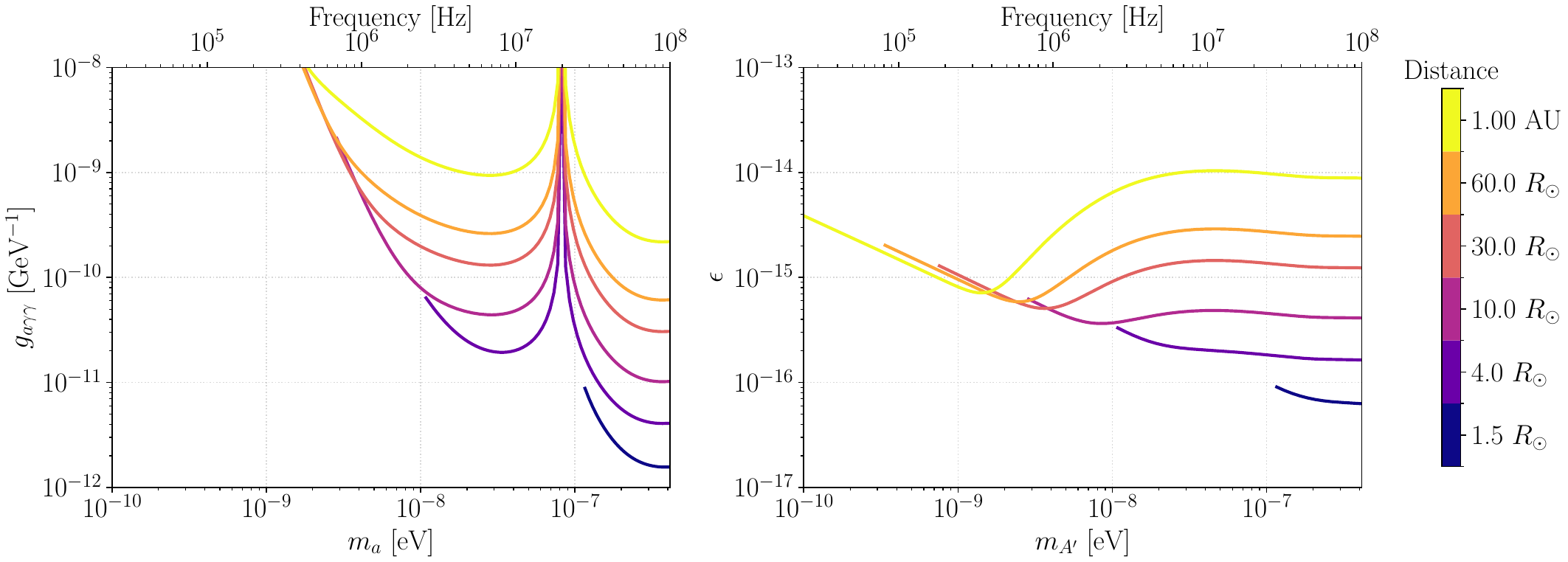}
  \caption{Sensitivity to DM ALP (left) and DP (right) radio lines from the Sun, for sky-dominated analyses (Galactic background plus QTN evaluated at the corresponding observation distance) and for different observation distances (colour bar). For DPs, the curves exhibit a turnover at the mass scale where the QTN contribution overtakes the Galactic background, which depends on the observation distance $d$}.
  
  \label{fig:sens_Gal_Sun}
\end{figure*}

\begin{figure*}[t] 
  \centering
  \includegraphics[width=2.1\columnwidth]{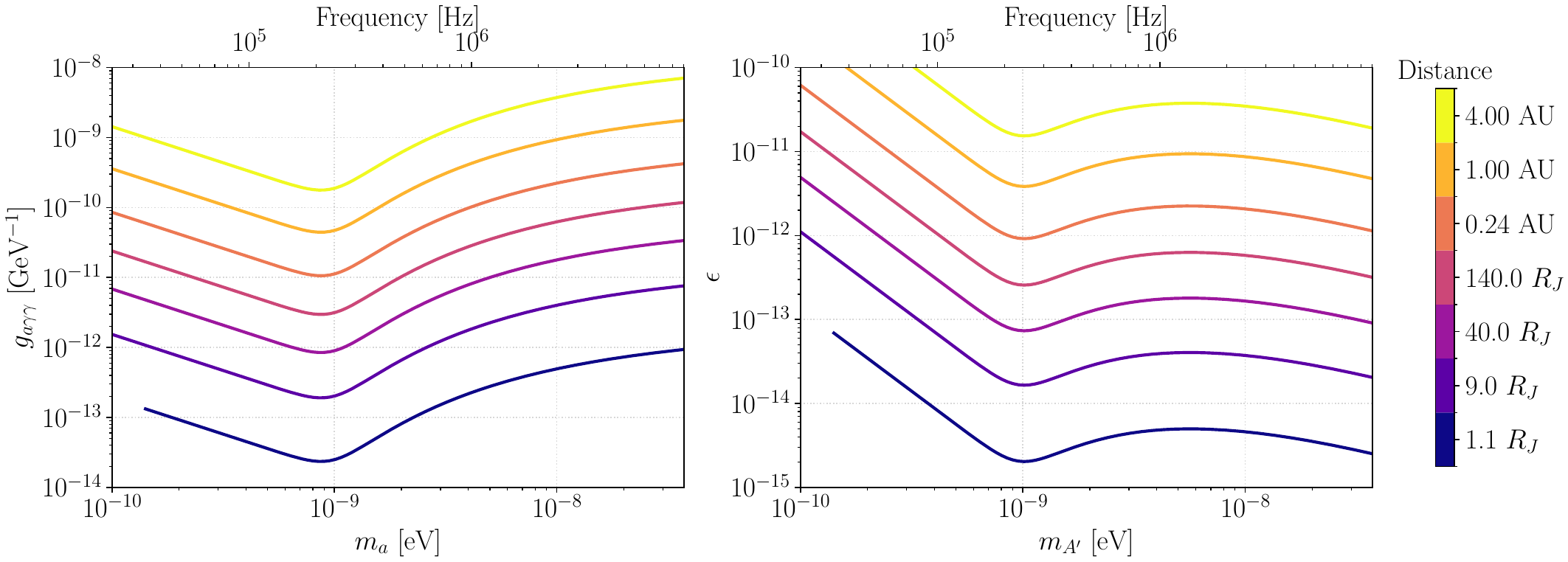}
  \caption{Sensitivity to DM ALP (left) and DP (right) radio lines from Jupiter, for sky-dominated analyses (Galactic background plus QTN at 5 AU, approximately the orbital distance of Jupiter) and for different observation distances (colour bar). The curves exhibit a clear turnover around $m_a\!\sim\!10^{-9}\,\mathrm{eV}$, corresponding to the mass below which the QTN contribution exceeds the Galactic background.}
  \label{fig:sens_Gal_Jupiter}
\end{figure*}

\begin{figure*}[t]
  \centering
  \includegraphics[width=2.1\columnwidth]{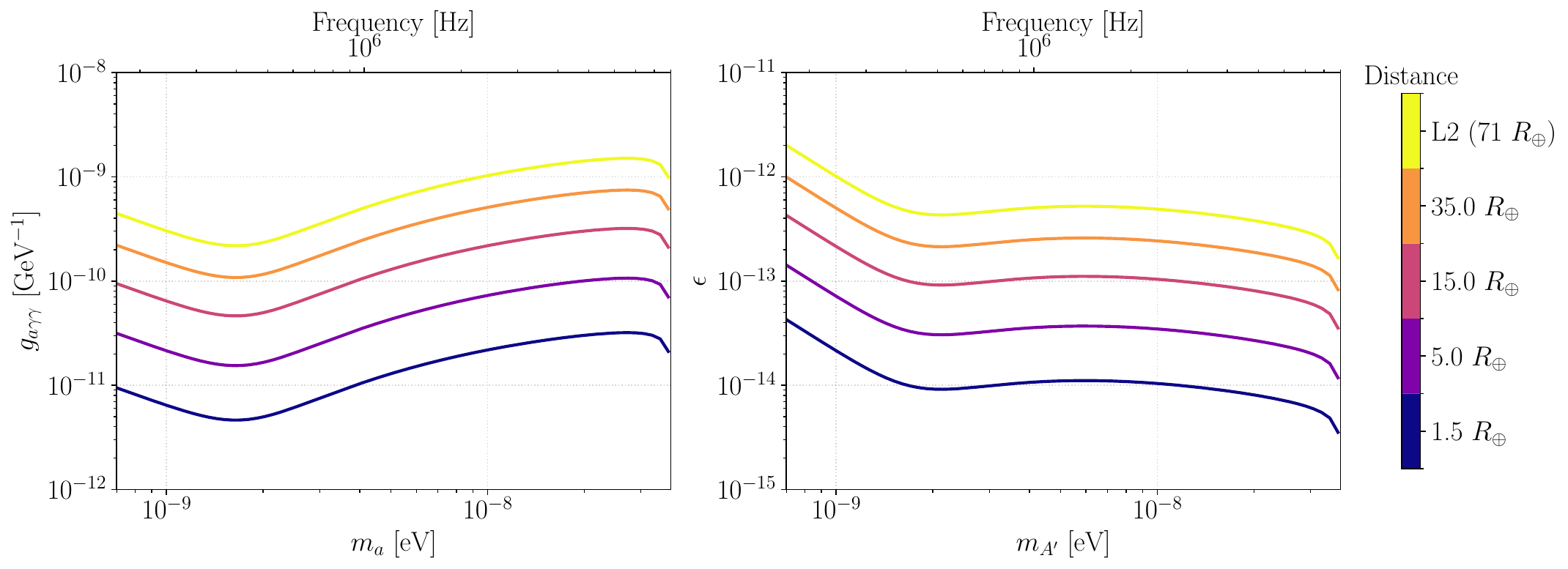}
  \caption{Sensitivity to DM ALP (left) and DP (right) radio lines from Earth, for sky-dominated analyses (Galactic background plus QTN at 1 AU, the orbital distance of Earth) and for different observation distances (colour bar). The curves exhibit a clear turnover around $m_a\!\sim\!10^{-9}\,\mathrm{eV}$, corresponding to the mass below which the QTN contribution exceeds the Galactic background.}
  \label{fig:sens_Gal_Earth}
\end{figure*}

Figs.~\ref{fig:sens_Gal_Sun}--\ref{fig:sens_Gal_Earth} show the resulting sensitivity projections.
In the solar case, an excellent sensitivity to  DP signals is reachable, in principle even below $\epsilon\simeq 10^{-16}$, while the detection of ALPs can barely surpass  $g_{a\gamma\gamma}\!\sim\!10^{-11}\,{\rm GeV^{-1}}$ at high masses ($m_a\gtrsim 10^{-7}$eV) and only for observation distances very close to the Sun.
Instead, monitoring Jupiter can allow one to test $g_{a\gamma\gamma}$ even below $g_{a\gamma\gamma}\!\sim\!10^{-12}\,{\rm GeV^{-1}}$, but is less sensitive to $\epsilon$ compared to the observation of the Sun. Note that the sensitivities at low masses are limited by the QTN, while at higher masses by the Galactic background. The two regimes are clearly seen in the plots referring to DP: the transition occurs at $m_{A'} \sim 2\times10^{-9}$ eV for the Earth and at $m_{A'} \sim 8\times10^{-10}$ for Jupiter, reflecting the weaker QTN level at 5 AU due to the lower local plasma density and temperature. For the Sun, the turnover in mass varies with distance because of the radial dependence of the QTN.

\subsection{Including instrumental noise}
We now include the receiver contribution to the total system noise. 
Instruments marked with an asterisk in Tab.\;\ref{tab:instrument_specs} already quote SEFDs that include the environmental contribution (Galactic and/or QTN), and these are used as provided.
For all others, we compute the SEFD by adding the environmental component to the receiver noise, in line with the modelling in Sec.~\ref{sec:detection}.

\begin{figure*}[t] 
  \centering
  \includegraphics[width=2\columnwidth]{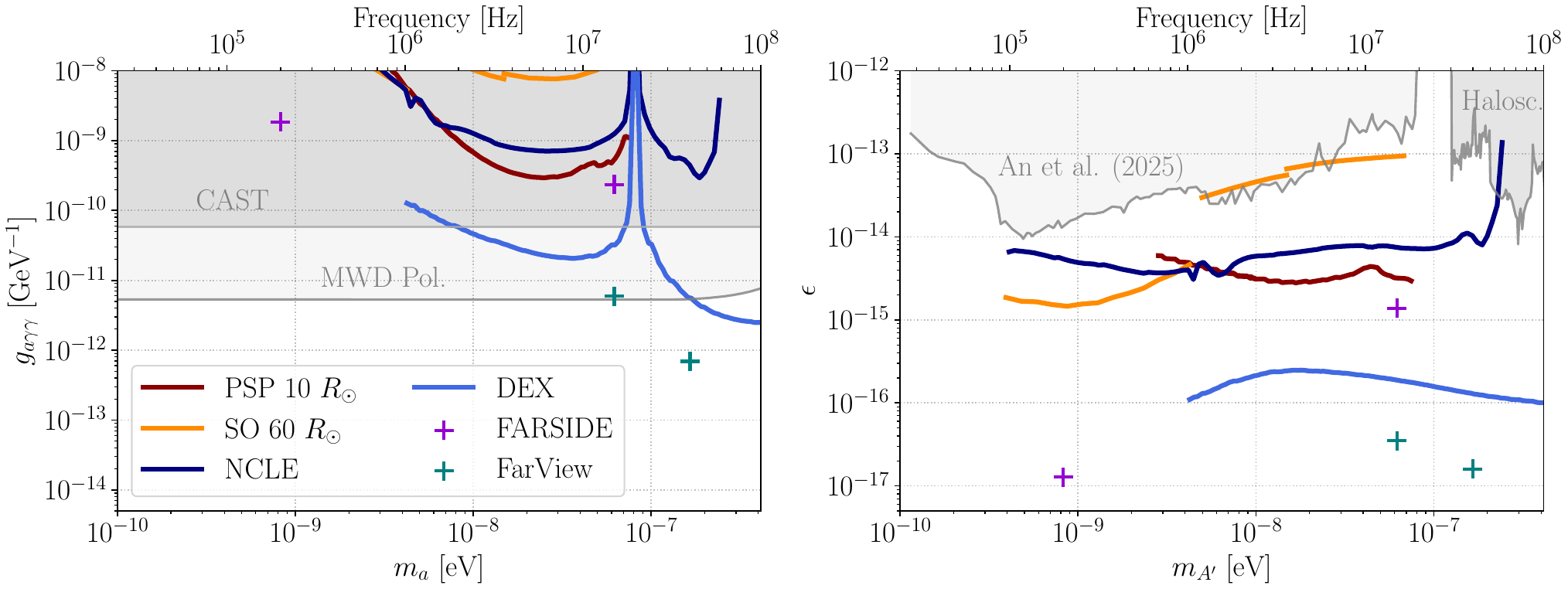}
  \caption{Sensitivity to DM ALP (left) and DP (right) line signals from the Sun for specific lunar and solar probes. For reference, we overlay
  also current constraints: On the left, the shaded dark grey area displays the current CAST limits~\cite{CAST:2024eil}, while the light grey one the bounds from magnetic white dwarfs (MWD) polarisation~\cite{Dessert:2022yqq}; on the right, the light grey region refers to the analysis of PSP by~\cite{2025PhRvL.134q1001A}, while the dark grey one to the haloscopes limits, according to  \url{https://cajohare.github.io/AxionLimits/docs/dp.html}.} 
  \label{fig:sens_instr_Sun}
\end{figure*}

\begin{figure*}[t] 
  \centering
  \includegraphics[width=2\columnwidth]{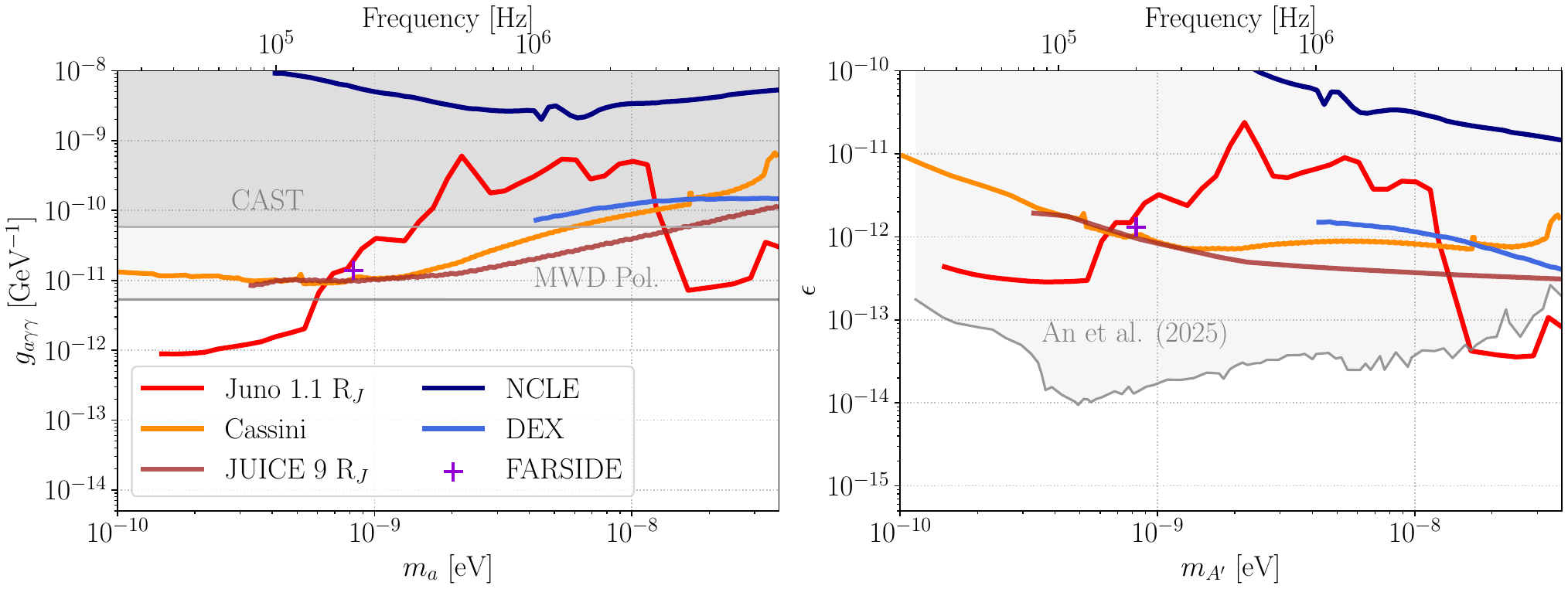}
  \caption{Sensitivity to DM ALP (left) and DP (right) line signals from the Sun for specific lunar and solar probes. For reference, we overlay
  also current constraints: On the left, the shaded dark grey area displays the current CAST limits~\cite{CAST:2024eil}, while the light grey one the bounds from magnetic white dwarfs (MWD) polarisation~\cite{Dessert:2022yqq}; on the right, the light grey region refers to the analysis of PSP by~\cite{2025PhRvL.134q1001A}, while the dark grey one to the haloscopes limits, according to \cite{Caputo:2021eaa}.}
  \label{fig:sens_instr_Jupiter}
\end{figure*}

\begin{figure*}[t] 
  \centering
  \includegraphics[width=\textwidth]{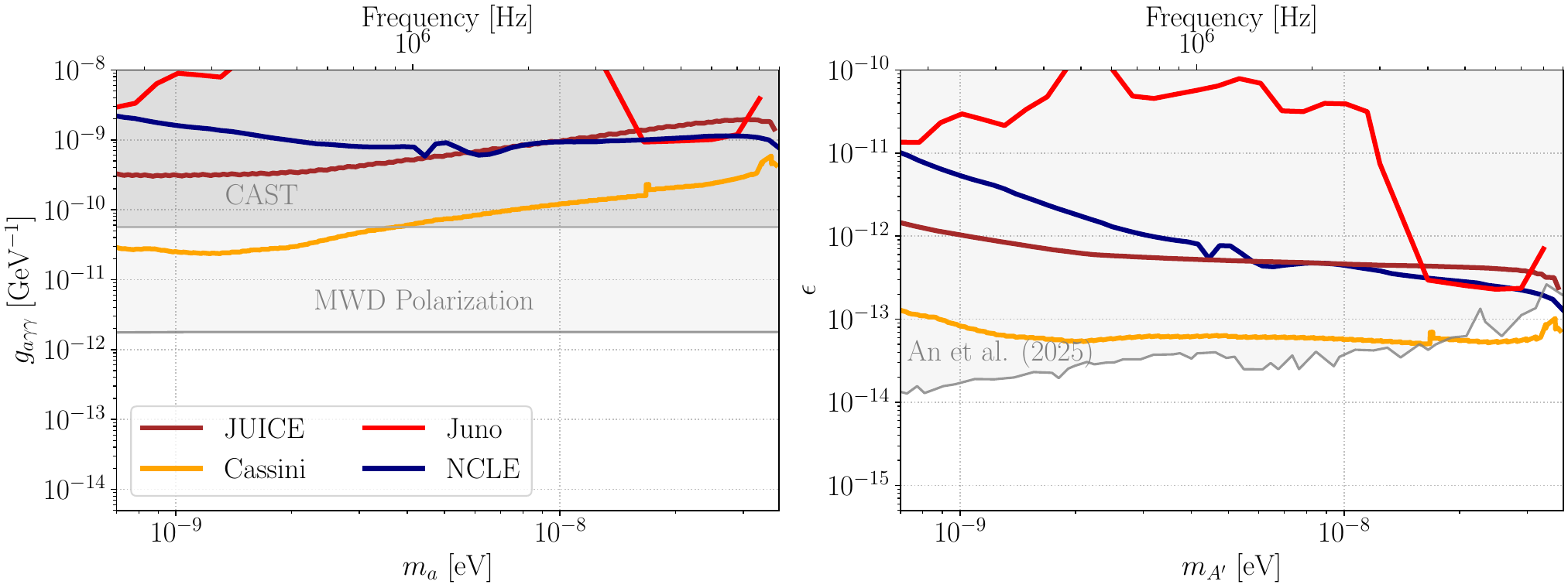}
  \caption{Sensitivity to DM ALP (left) and DP (right) line signals from the Sun for specific lunar and solar probes. For reference, we overlay
  also current constraints: On the left, the shaded dark grey area displays the current CAST limits~\cite{CAST:2024eil}, while the light grey one the bounds from magnetic white dwarfs (MWD) polarisation~\cite{Dessert:2022yqq}; on the right, the light grey region refers to the analysis of PSP by~\cite{2025PhRvL.134q1001A}, while the dark grey one to the haloscopes limits, according to \cite{Caputo:2021eaa}.}
  \label{fig:sens_instr_Earth}
\end{figure*}

The resulting sensitivity curves for ALPs and DPs are presented in Figs.~\ref{fig:sens_instr_Sun}--\ref{fig:sens_instr_Earth}. 

As already observed in the environmental noise-only analysis, the Sun remains the most promising target for probing the kinetic mixing coupling $\epsilon$. In particular, future lunar–farside interferometric arrays (DEX, FARSIDE, FarView) show a remarkable sensitivity due to the large number of antennas, accessing unexplored regions of the 
parameter space, reaching $\epsilon \sim 10^{-14}$--$10^{-17}$ for $m_{A'} \sim 3 \times 10^{-10}$--$5 \times 10^{-7}$ eV.
On the other hand, the reach in the ALP-photon coupling $g_{a\gamma\gamma}$ is more limited, but some of lunar arrays
can probe couplings below the CAST experiment (\(g_{a\gamma\gamma}\lesssim 6 \times 10^{-11}\,{\rm GeV^{-1}}\))~\cite{CAST:2024eil}.
Between the two solar probes, PSP outperforms SO thanks to its closer proximity to the Sun and its lower receiver noise. 
We notice that the mass range probed by PSP is narrower than then instrument nominal frequency interval, because of 
the cut due to $d_{\rm min} = R_c(m_\alpha)$, as explained earlier.

For Jupiter, the situation is reversed, as the ALP scenario is more favourable. The Juno, JUICE and Cassini spacecrafts can reach values of $g_{a\gamma\gamma}$ well below benchmark bounds such as those from CAST, potentially down to $g_{a\gamma\gamma}\sim10^{-12}\,{\rm GeV^{-1}}$ in the case of Juno’s observations. 
Both the Juno and JUICE instruments are dominated by intrinsic receiver noise, preventing them from fully exploiting their short distance from Jupiter (see Fig.~\ref{fig:sens_Gal_Jupiter} for comparison at equal $d$). The Cassini spacecraft, on the other hand, has high instrumental sensitivity, but the larger distance of the flyby strongly suppresses the detection potential.

Despite the proximity achieved during spacecrafts' flybys, the Earth is a weaker target for ALP and DP detection due to the intrinsically faint DM signal. The Juno and JUICE curves show the limitations due to the high instrumental noise, while Cassini, thanks to its better sensitivity, can probe lower values of both $g_{a\gamma\gamma}$ (down to \(\sim 3 \times 10^{-11}\,{\rm GeV^{-1}}\)) and $\epsilon$ (down to \(\sim 4\)--5 \(\times 10^{-14}\)). 

\section{Discussions and conclusions}\label{sec:conclusions}
In this work, for the first time, we have systematically assessed the sensitivity 
to DM-photon conversion in nearby, solar system objects: the Sun, Jupiter and the Earth.
We adapted the generic formalism of resonant conversion in unmagnetised plasmas to the case of 
two specific light DM models, i.e.~ALPs and DPs, and evaluated the 
sensitivity of past and future space radio probes to these signals.
In projecting sensitivities, we considered both environmental radio emission from the Galactic  
background and QTN from the local plasma, as well as instrumental noise. 

Our analysis of those three resonant–conversion targets demonstrated that radio observations from space can probe previously unexplored regions of ALP and DP DM parameter space. Despite the presence of irreducible environmental backgrounds, meaningful reach remains attainable, in particular for DPs with observations of the Sun and,  to a lesser extent, for ALPs with observations of Jupiter. The finding that Jupiter outperforms the Sun for ALP resonant conversion is an a priori non-trivial outcome of the different magnetic and plasma scale lengths characterising the two environments.

Current and past spacecrafts around Jupiter, although not built for this specific science case, already show sensitivities to $g_{a\gamma\gamma}$ below leading CAST benchmark limit $(g_{a\gamma\gamma}\lesssim 6 \times 10^{-11}\,{\rm GeV^{-1}}$)~\cite{CAST:2024eil}. They are also competitive with the most aggressive limits in the literature ($g_{a\gamma \gamma} \lesssim 5.4 \times 10^{-12} \, \rm GeV^{-1}$)~\cite{Dessert:2022yqq}, which suffer, on their side, from much larger uncertainties in the astrophysical parameters of magnetic white dwarfs.
For DPs, the Cassini Earth flyby reaches $\epsilon$ values comparable with the strongest existing constraints\;\cite{2025PhRvL.134q1001A}, while solar observations—especially with future lunar farside interferometric arrays—offer the best prospects, potentially close to $\epsilon \sim 10^{-17}$.

More quantitatively, combining Eq.~\eqref{eq:average_SFD} with the radiometer condition Eq.~\eqref{eq:radiometer}, we infer that:
\begin{equation}
\label{eq:gmin_scaling}
    g_i^{\rm min} \propto
    \left[
    {\rm SNR}\,\cdot\,{\rm SEFD}
    \right]^{1/2}
    \, d \,
    \left(\frac{\Delta\nu}{t_{\rm int}}\right)^{1/4}.
\end{equation}
The sensitivity therefore improves linearly with decreasing distance to the target, making $d$ the most effective parameter to optimise. In-situ spacecrafts benefit from this scaling, achieving good sensitivity even when affected by elevated receiver noise, as in the case of Juno and JUICE. For these missions, the distance $d$ is already minimised, but a similar experiment equipped with a less noisy receiver (i.e., with an improved $\mathrm{SEFD_{\rm rx}}$) could reach the optimal sensitivity curves shown at the bottom of Fig.~\ref{fig:sens_Gal_Jupiter}, improving the limits by at least an order of magnitude across the relevant mass range.
We recall that the SEFD depends on the number of antennas so that $g_i^{\rm min} \propto N_{\rm ant}^{-1/2}$, explaining the excellent performance of interferometric arrays. This effect is particularly relevant for observations of DP converting in the Sun. 
Finally, the detectability improves with increasing integration time and decreasing bandwidth, although the dependence is weak. However, at low frequencies, current instruments typically operate with fractional spectral resolutions of order $10^{-2}$, while the intrinsic DM line width is expected to be $\Delta\nu_{\rm line}/\nu_\alpha \sim 10^{-6}$. Therefore, in an ideal experiment capable of resolving the full line width, the sensitivity could improve by an order of magnitude in low frequency range ($< 1\, \rm{MHz}$). Moreover, we have seen how spacecrafts performing Earth flybys typically remain within the optimal observation region for at most $\sim1$~hour. Extending the observing time to $\sim100$~hours would yield an improvement in the corresponding limits by a factor of about three.

The analysis presented here focuses on the irreducible (stationary) contributions to the system noise that set a fundamental sensitivity floor. Time-variable (non-stationary) radio emission from the targets themselves (e.g.\ solar or planetary bursts) cannot be treated within the simple radiometer–equation framework and must instead be identified and removed, or explicitly modelled, in a dedicated data analysis.

In this context, it is instructive to refer to the DP search performed with PSP data in Ref.~\cite{2025PhRvL.134q1001A}, to  be compared with our PSP forecast in Fig.\;\ref{fig:sens_instr_Sun}. In that work, the authors first select relatively quiet time–frequency intervals (i.e.\ by retaining only the lowest 3\% of data points in each bin), then fit a smooth background spectrum in each frequency bin and construct a likelihood profile to test for the extra DP narrow line. The corresponding limits on $\epsilon$ are quoted at 95\%~C.L.
By contrast, our idealised sensitivity estimate---based on a constant distance equal to the minimum perihelion, $d = 10\,R_{\odot}$, an effective integration time $t_{\rm int}=100$~hr, and a SEFD that includes only instrumental, Galactic, and QTN noise---yields sensitivities to $\epsilon$ that are more than an order of magnitude stronger than those derived from the real PSP data in Ref.~\cite{2025PhRvL.134q1001A}. This discrepancy can be traced back to several effects:
(i) the average Sun--spacecraft distance in the real dataset are less favourable than in our simplified setup 
and
(ii) residual solar and instrumental variability of the data are not fully captured by the smooth background model (as indicated by reduced $\chi^2$ values always larger than unity in Fig.~S3 of Ref.~\cite{2025PhRvL.134q1001A}). 
Our curves should therefore be interpreted as optimistic benchmarks, illustrating the 
 reach of space-based telescope with ideal noise properties.

To conclude, we have presented a first systematic, though simplified, approach to study the potential of solar system targets with present and up-coming observations. Our encouraging results suggest that these searches constitute an additional scientific case for future space-based or lunar-based radio-surveys; they deserve follow-up dedicated studies and analyses, motivating  more realistic and mission-specific assessments.

\bigskip
\begin{acknowledgments}
FC thanks M.~Regis for fruitful discussions.
The work of FC, PDS and RZ is supported by the European Union through the grant ``UNDARK'' of the Widening participation and spreading excellence programme (project number 101159929). RZ also acknowledges the MICINN through the grant ``DarkMaps'' PID2022-142142NB-I00.
\end{acknowledgments}

\bigskip
\section*{data availability}
The data that support the findings of this article are openly available \cite{zatini2026lowfreqradiodm}.

\bibliographystyle{bibi}
\bibliography{references}

\end{document}